\documentclass[nofootinbib]{revtex4}

\usepackage[english]{babel}
\usepackage{graphicx}
\usepackage{psfrag}
\usepackage{amsmath}
\usepackage{amssymb}
\usepackage{verbatim}

\newcommand{\be}{\begin{equation}}
\newcommand{\ee}{\end{equation}}
\newcommand{\bea}{\begin{eqnarray}}
\newcommand{\eea}{\end{eqnarray}}
\newcommand{\bean}{\begin{eqnarray*}}
\newcommand{\eean}{\end{eqnarray*}}

\renewcommand{\b}{\langle}
\newcommand{\ket}{\rangle}

\newcommand{\irm}{{\rm i}}
\newcommand{\e}{{\rm e}}

\renewcommand{\d}{{\rm d}}
\newcommand{\cl}[1]{{\mathcal #1}}

\newcommand{\pa}{\partial}

\newcommand{\ts}{\textstyle}

\newcommand{\bZ}{\mathbb{Z}}

\newcommand{\bR}{\mathbb{R}}

\newcommand{\clN}{\cl{N}}

\newcommand{\clS}{\cl{S}}

\newcommand{\eq}[1]{(\ref{#1})}

\newcommand{\tr}{{\rm tr}}

\newcommand{\qed}{\nobreak \ifvmode \relax \else
      \ifdim\lastskip<1.5em \hskip-\lastskip
      \hskip1.5em plus0em minus0.5em \fi \nobreak
      \vrule height0.75em width0.5em depth0.25em\fi}

\newcommand{\muh}{\hat{\mu}}
\newcommand{\nuh}{\hat{\nu}}
\newcommand{\oneh}{\hat{1}}

\newcommand{\bb}{\overline{b}}

\newcommand{\nablab}{\overline{\nabla}}

\begin{document}
\thispagestyle{empty}
\hfill
\parbox[t]{3.6cm}{
hep-th/yymmnnn \\
IGPG-06/10-4} 
\vspace{2cm}

\title{Analytic derivation of dual gluons and monopoles \\ from SU(2) lattice Yang-Mills theory \\ I.\ BF Yang-Mills representation}
\author{Florian Conrady}
\affiliation{Institute for Gravitational Physics and Geometry, Physics Department, Penn State University, University Park, PA 16802, U.S.A}
\email{conrady@gravity.psu.edu}
\preprint{IGPG-06/10-4}

\begin{abstract}
In this series of three papers, we generalize the derivation of dual photons and monopoles by Polyakov, and Banks, Myerson and Kogut,
to obtain approximative models of SU(2) lattice gauge theory. The papers take three different representations as their starting points: the representation as a BF Yang-Mills theory, the spin foam representation and the plaquette representation. The derivations are based on stationary phase approximations.

In this first article, we cast 3- and 4-dimensional SU(2) lattice gauge theory in the form of a lattice BF Yang-Mills theory. In several steps, the expectation value of a Wilson loop is transformed into a path integral over a dual gluon field and monopole-like degrees of freedom. The action contains the tree-level Coulomb interaction and a nonlinear coupling between dual gluons, monopoles and current.
\end{abstract}

\maketitle

\section{Introduction}
\label{introduction}

The analysis of QCD and its low-energy physics is one of the major challenges of present-day theoretical physics (for a review, see e.g.\ \cite{Greensitereview}). The main difficulty lies in the fact that many of the relevant phenomena happen at distance scales where the effective coupling is large and perturbative techniques cease to be applicable. Thus, it becomes necessary to devise non-perturbative methods that can predict the effective physics at these scales.

So far lattice simulations are the most successful tool in this regime: they provide a wealth of data, and become more and more accurate as computation power increases. The drawback is that the data in itself do not explain the underlying physical mechanisms. Therefore, it is also essential to have analytic derivations that produce models of confinement, chiral symmetry breaking and other phenomena, and can be compared with the lattice data.

As far as confinement is concerned, we have the following well-established analytic results: 
in the strong-coupling regime of non-abelian lattice gauge theory, the area law was demonstrated by expansions in strong-coupling graphs\footnote{This does not constitute a satisfactory proof, however, since it requires a cutoff scale that is comparable to the confinement scale.}  \cite{Wilsonconfinement,KogutPearsonShigemitsu,Munsterhightemperature,DrouffeZuber}. For U(1) lattice gauge theory in 3 dimensions confinement has been derived by Polyakov \cite{PolyakovI,PolyakovII} and Banks, Myerson and Kogut \cite{BanksMyersonKogut}, and rigorously proven by G\"{o}pfert and Mack \cite{GopfertMack}. In 4 dimensions, one has a phase transition between a confining and a deconfined phase, which was shown by Banks et al., Guth, and Fr\"{o}hlich \& Spencer \cite{BanksMyersonKogut,Guth,FrohlichSpencer}.

There are two main approaches to go beyond these results: a string-theoretic and a field-theoretic one. 
A main motivation for the string-theoretic approach is the fact that, for strong coupling, the confining potential is produced by electric flux lines (or strings) between quarks\footnote{The aforementioned strong-coupling graphs are the worldsheets of these electric flux lines.}. One hopes to find a string representation of Yang-Mills theory that could explain confinement in the continuum limit (for a review, see \cite{Polyakovgaugefieldsstrings,Polyakovliberation,Polyakovconfiningstrings,Antonovstringnature}). In this regard, progress was made by establishing correspondences between superstring theories and super Yang-Mills theories on AdS spacetimes \cite{Maldacena}. 

The field-theoretic strategy is motivated by the example of U(1), where confinement results from monopole condensation between charges. 
The aim is to generalize this to non-abelian gauge groups and explain confinement as an effect of special types of gauge-field configurations (for a review, see \cite{Greensitereview} and \cite{Polyakovliberation}): candidates are, for example, monopoles in the maximal abelian gauge \cite{tHooftmaximalabeliangauge,ChernodubPolikarpov}, center vortices, and instantons. So far we do not know of a direct analytic way to derive the effective actions for these objects. In the maximal abelian gauge, one obtains an abelian gauge theory that contains more than just photons and monopoles, and its analysis is complicated. An important result is that by Seiberg and Witten on monopoles in supersymmetric Yang-Mills theory \cite{SeibergWittenelectric,SeibergWittenmonopoles}. It is not clear, however, if it is of direct relevance for non-abelian confinement (see e.g.\ \cite{Kovner}). Further proposals for monopole actions can be found in refs.\ \cite{SmitSijs,DasWadia,Gromes}.

There exist also novel approaches that do not belong to the two categories we just described: for example, for 3 dimensions, Karabali, Kim \& Nair have developed a strong-coupling expansion that does not require any lattice regularization and yields an analytic derivation of the string tension \cite{KarabaliNair96,KarabaliKimNair98,KarabaliKimNair2000}. Orland obtained the confining potential in a certain weak-coupling limit with anisotropic couplings \cite{Orlandintegrable,Orlandstringtensions,Orlandconfinementweakcoupling}. Leigh, Minic \& Yelnikov derived analytic results in the large $N$ limit \cite{LeighMinicYelnikovsolving2plus1,LeighMinicYelnikovhowtosolve2plus1}. 

In this series of papers, we develop a new approach to the monopole model of confinement, and propose analytic derivations of gluon-monopole actions for SU(2) lattice Yang-Mills theory. The three papers take three different representations as their starting points: the BF Yang-Mills representation in dimension 3 and 4, the spin foam representation in $d=3$ \cite{ConradyglumonII}, and the plaquette representation in $d=3$ \cite{ConradyglumonIII}. In each case, we approximate the expectation value of a Wilson loop by a path integral over a dual gluon field and monopole-like excitations. 

The resulting gluon-monopole actions are not identical, but similar: in all three cases, the tree-level Coulomb interaction is roughly reproduced, and the coupling between monopoles, dual gluons and current resembles that of the abelian case. There is an important difference, however: it consists in the fact that the dual gluon (or Debye-H\"{u}ckel) field is $\mathrm{su(2)}\simeq \bR^3$-valued (and not $\bR$-valued) and that the monopoles couple to the \textit{length} of field vectors. This renders the gluon-monopole coupling nonlinear.

In this paper, we start from the BF Yang-Mills representation in dimension 3 and 4.
By this we mean that we cast the SU(2) lattice gauge theory in a form that can be regarded as a lattice version of BF Yang-Mills theory \cite{MartelliniZeni}.
Due to the compactness of SU(2), the lengths $|B|$ of the $B$-field are restricted to discrete half-integer values. 
By applying the Poisson summation formula, we can trade this discreteness for a continuous variable and a discrete monopole variable. After making a stationary phase approximation, we obtain a constraint that is analogous to the abelian Gauss constraint. Solving the constraint yields the dual gluon degrees of freedom. 

The paper is organized as follows: 
in sec.\ \ref{SU2latticeYangMillstheory}, we set the conventions for SU(2) lattice gauge theory and briefly review its representations.
Section \ref{SU2latticeYangMillstheoryasalatticeBFYangMillstheory} describes the rewriting as a lattice BF Yang-Mills theory. 
The main result is presented in section \ref{representationasgluonsandmonopolelikeexcitations}, where we derive the representation in terms of dual gluons and monopole-like excitations. In the final section, we summarize and discuss the results.
 
\subsubsection*{\textit{Notation and conventions}}

$\kappa$ denotes a $d$-dimensional hypercubic lattice of side length $L$ with periodic boundary conditions. The lattice constant is $a$. 
Depending on the context, we use abstract or index notation to denote oriented cells of $\kappa$: in the abstract notation, vertices, edges, faces and cubes are written as $v$, $e$, $f$ and $c$ respectively. In the index notation, we write $x$, $(x\mu)$, $(x\mu\nu)$, $(x\mu\nu\rho)$ etc. Correspondingly, we have two notations for chains. Since the lattice is finite, we can identify chains and cochains. As usual, $\pa$, $\d$ and $*$ designate the boundary, coboundary and Hodge dual operator respectively. Forward and backward derivative are defined by
\be
\nabla_\mu f_x = \frac{1}{a}\left(f_{x+a\muh} - f_x\right)\,,\qquad \nablab_\mu f_x = \frac{1}{a}\left(f_x - f_{x-a\muh}\right)
\ee
where $\muh$ is the unit vector in the $\mu$-direction. The lattice Laplacian reads
\be
\Delta = \nablab_\mu \nabla_\mu\,.
\ee
For a given unit vector $u = \muh$ and a 1-chain $J_{x\mu}$, we define
\be
(u\cdot\nabla)^{-1} J_{x\mu} := \sum_{x'_\mu \le x_\mu} J_{(x_1,\ldots,x'_\mu,\ldots,x_d)\,\mu}\,.
\ee
Color indices are denoted by $a, b, c, \ldots$ We employ units in which $\hbar = c = 1$ and $a = 1$. For some quantities, the $a$-dependence is indicated explicitly.

\section{SU(2) lattice Yang-Mills theory}
\label{SU2latticeYangMillstheory}

In this section, we set our conventions for SU(2) lattice Yang-Mills theory, and briefly describe the present knowledge about its representations.
For the properties of the lattice $\kappa$, see the end of the introduction. For the moment, we assume that $d\ge 2$, but later we will consider only dimension 3 and 4.

The partition function of $d$--dimensional SU(2) lattice gauge theory is defined by a path integral over SU(2)--valued edge (or link) variables $U_{x\mu}$:
\be
\label{definingrepresentationSUtwo}
Z = \int\left({\ts\prod\limits_{x\mu}}\;\d U_{x\mu}\right) \exp\left[-\sum_x\sum_{\mu<\nu} \clS_{x\mu\nu}(W_{x\mu\nu})\right]
\ee
The plaquette (or face) action $\clS_{x\mu\nu}$ depends on the holonomy
\be
W_{x\mu\nu} = U^{-1}_{x,\nu}U^{-1}_{x+a\nuh,\mu}U_{x+a\muh,\nu}U_{x\mu}
\ee
around the plaquette. Here, we will use the heat kernel action \cite{MenottiOnofri}. Let us write group elements as exponentiations of Lie algebra elements, i.e.
\be
\label{plaquetteholonomy}
U_{x\mu} = \e^{\irm\,\theta^a_{x\mu} \sigma^a/2}\,,\qquad\mbox{and}\qquad W_{x\mu\nu} = \e^{\irm\,\omega^a_{x\mu\nu}\sigma^a/2}\,,
\ee
where $\sigma^a$, $a = 1,2,3$, are the Pauli matrices and $|\theta_{x\mu}|, |\omega_{x\mu\nu}|<2\pi$. Then, the heat kernel action is given by
\be
\label{heatkernelaction}
\exp\left(-\clS_{x\mu\nu}(W_{x\mu\nu})\right)\; =\;
\clN\sum_{n\in\bZ}\,\frac{|\omega_{x\mu\nu}| + 4\pi n}{\sin(|\omega_{x\mu\nu}|/2)}\,\exp\left(-\frac{1}{8}\,\beta\left(|\omega_{x\mu\nu}| + 4\pi n\right)^2\right)
\ee
The coupling factor $\beta$ is related to the gauge coupling $g$ via
\be
\beta = \frac{4}{a^{4-d} g^2} + \frac{1}{3}\,.
\ee
The SU(2) lattice gauge theory can be cast into equivalent representations that involve different degrees of freedom. We presently know of three such representations: a first-order representation, which can be viewed as a lattice version of BF Yang-Mills theory,
the spin foam representation and the plaquette representation. The first two representations exist in any dimension $d\ge 2$, while the latter has been only constructed in 3 dimensions so far. The plaquette representation is obtained by taking the holonomies $W_{x\mu\nu}$ around faces as the basic variables, subject to a non-abelian counterpart of the Bianchi constraint \cite{Batrouni,BatrouniHalpern,BorisenkoVoloshinFaber}. The first-order representation results from an expansion of plaquette actions into characters and has two sets of variables: the original edge variables $U_{x\mu}$, and spin assignments $j_{x\mu\nu}$ to plaquettes. As in U(1) lattice gauge theory, we can perform an exact integration over the edge variables: it yields a sum over so-called spin foams---configurations that consist of assignments of spins $j_{x\mu\nu}$ and intertwiners $I_{x\mu}$ to plaquettes and edges respectively\footnote{Spin foams are essentially the same as the strong-coupling graphs of the strong-coupling expansion \cite{Wilsonconfinement,Munsterhightemperature,KogutPearsonShigemitsu,DrouffeZuber}. Their appearance, however, is not tied to any expansion in the coupling, so we prefer to use the term ``spin foam'', which was coined in the quantum gravity literature \cite{Baezspinfoammodels}. For a review of gravity spin foams, see e.g.\ \cite{Perezreview}.} \cite{Anishettyetal,HallidaySuranyi,OecklPfeifferdualofpurenonAbelian,Conradygeometricspinfoams}.


\section{SU(2) lattice Yang-Mills theory as a lattice BF Yang-Mills theory}
\label{SU2latticeYangMillstheoryasalatticeBFYangMillstheory}

In the present paper, we start from a representation of lattice gauge theory that we call the BF Yang-Mills representation.
At the classical level, BF Yang-Mills theory is a certain deformation of BF theory and equivalent to Yang-Mills theory.
It has been demonstrated that BFYM and YM theory are equivalent at the quantum level, when quantized perturbatively in the continuum \cite{MartelliniZeni,Cattaneoetal,Accardietal1,Accardietal2}. 

With the help of the Kirillov trace formula, we can relate these theories also non-perturba\-tive\-ly on the lattice. 
To our knowledge, this has not been pointed out in the literature so far, so we explain it in this section.

We want to consider the low-coupling regime, that is, $\beta\gg 1$. After expanding the plaquette actions into characters, the partition function \eq{definingrepresentationSUtwo} takes the following form:
\be
\label{firstorder}
Z = \int\left({\ts\prod\limits_{x\mu}}\;\d U_{x\mu}\right) \sum_{j_{x\mu\nu}}\;
\left({\ts\prod\limits_{x\mu\nu}}\;(2j_{x\mu\nu}+1)\,\chi_{j_{x\mu\nu}}(W_{x\mu\nu})\,\,\e^{-\frac{2}{\beta}\,j_{x\mu\nu}(j_{x\mu\nu} + 1)}\right)
\ee 
By using the Kirillov trace formula \cite{Kirillov}, we can rewrite \eq{firstorder} in such a way that it appears 
like a BF Yang-Mills theory on a lattice. According to the Kirillov trace formula, the character is equal to an integral over unit vectors $n$ in $\bR^3$:
\be
\chi_j\left(W_{x\mu\nu}\right) = 
\frac{(2j+1)|\omega_{x\mu\nu}|/2}{4\pi\sin(|\omega_{x\mu\nu}|/2)}\,\int_{S^2} \d n\;\,\e^{\irm\,(2j+1)\,n\cdot\omega_{x\mu\nu}/2}
\ee
Since $\beta$ is large, spins are only weakly damped in \eq{firstorder}. Therefore, spins are typically large, and we use the approximations
\be
2j_{x\mu\nu}+1 \approx 2j_{x\mu\nu}\,,\qquad j_{x\mu\nu}\;(j_{x\mu\nu} + 1) \approx j^2_{x\mu\nu}\,.
\ee
We also omit any field-independent factors that would drop out in expectation values. Thus, we get
\bea
\lefteqn{Z = \int\left({\ts\prod\limits_{x\mu}}\;\d U_{x\mu}\right)\;
\sum_{j_{x\mu\nu}} \left({\ts\prod\limits_{x\mu\nu}} j_{x\mu\nu}\right)\;
\int_{S^2}\left({\ts\prod\limits_{x\mu\nu}}\d n_{x\mu\nu}\right)} \nonumber \\
&& \times\,\left({\ts\prod\limits_{x\mu\nu}}\,\frac{|\omega_{x\mu\nu}|/2}{\sin(|\omega_{x\mu\nu}|/2)}\right)
\exp\left[\sum_{x\mu\nu}
\left(\frac{\irm}{2}\,j_{x\mu\nu} n_{x\mu\nu}\cdot \omega_{x\mu\nu} 
- \frac{1}{\beta}\, j^2_{x\mu\nu}\right)
\right]\,.
\eea 
We can think of the sum over $j$'s and integrals over 
$n$'s as an integral over vectors
\be
b_{x\mu\nu} = j_{x\mu\nu} n_{x\mu\nu}
\ee
in su(2) whose length is restricted to half-integer values. The path integral becomes
\bea
\lefteqn{Z = \int\left({\ts\prod\limits_{x\mu}}\;\d U_{x\mu}\right) 
\int_{\bR^3}\left({\ts\prod\limits_{x\mu\nu}}\;\d^3 b_{x\mu\nu}\,\sum_j\,\delta(|b_{x\mu\nu}| - j)\right)} \nonumber \\
&& \times\,\left({\ts\prod\limits_{x\mu\nu}}\,\frac{|\omega_{x\mu\nu}|/2}{\sin(|\omega_{x\mu\nu}|/2)}\right)
\exp\left[\sum_x
\left(\frac{\irm}{2}\,b_{x\mu\nu}\cdot\omega_{x\mu\nu} 
- \frac{1}{\beta}\, b^2_{x\mu\nu}\right)
\right]\,.
\label{latticeBFYM}
\eea
In the exponent repeated indices are summed over.

A gauge transformation $\lambda$ on the connection implies a rotation of the Lie algebra element $\omega_{x\mu\nu}$
and can be compensated by a corresponding rotation of $b_{x\mu\nu}$:
\be
\omega'{}^a_{x\mu\nu} = R_1(\lambda_x)^a{}_c\,\omega^c_{x\mu\nu}\,,\qquad b'{}^a_{x\mu\nu} = (R^{-1}_1(\lambda_x))^a{}_c\,b^c_{x\mu\nu}
\ee
$R_1$ stands for the adjoint representation. Therefore, in the representation \eq{latticeBFYM}, we can view gauge symmetry as a symmetry that involves both the connection and the $b$-field. If we define the dual form 
\be
b_{x\rho_1\cdots \rho_{d-2}} = \frac{1}{2}\,\epsilon_{\rho_1\cdots \rho_{d-2}\mu\nu}\,b_{x\mu\nu}\,,
\ee
the action takes the form
\be
\clS = \sum_x 
\left(
- \frac{\irm}{2!(d-2)!}\,\epsilon_{\mu\nu\rho_1\cdots \rho_{d-2}} b_{x\rho_1\cdots \rho_{d-2}}\cdot\omega_{x\mu\nu} 
+ \frac{2}{(d-2)!\,\beta}\, b^2_{x\rho_1\cdots \rho_{d-2}}
\right)
\ee
In the naive continuum limit, we have
\be
\omega = a^2 g\,F + o(a^3)\,,\qquad b = a^{d-2} g^{-1}\, B\,,
\ee
and the action approaches
\be
\clS = \sum_x\,a^d\,\frac{1}{2!(d-2)!}
\left(- \irm\,\epsilon_{\mu\nu\rho_1\cdots \rho_{d-2}} B_{x\rho_1\cdots \rho_{d-2}}\cdot F_{x\mu\nu} 
+ B^2_{x\rho_1\cdots \rho_{d-2}}\right)\,,
\ee
a discrete version of the continuum action of BF Yang-Mills theory. This shows that in the representation \eq{latticeBFYM} the lattice Yang-Mills theory can be viewed as a lattice version of BF Yang-Mills theory, where the lengths of the $B$'s are restricted to be discrete. The discreteness of lengths arises from the discrete set of character functions, and that is, in turn, a consequence of the compactness of SU(2). So altogether the compactness of SU(2) manifests itself in two ways in \eq{latticeBFYM}: the compact range of the group variables $U_{x\mu}$, and the discrete lengths of the $b_{x\mu\nu}$-variables.

Let us now introduce a source. Consider a Wilson loop $C$ in the representation $j$. We choose an arbitrary starting point $x_0$ in the Wilson loop and order the edges of $\kappa$ that coincide with it, following the orientation of the loop: $e_1 = (x_1,\mu_1)$, \ldots, $e_n = (x_n,\mu_n)$. The holonomy around $C$ is given by
\be
W_C = \left({\ts\prod\limits_{i=1}^n}\,U^{s_i}_{x_i\mu_i}\right) = U^{s_n}_{x_n\mu_n}\cdots U^{s_1}_{x_1\mu_1}\,,
\ee
where $s_i = 1$ if $e_i$ goes in the direction of $C$ and otherwise $s_i = -1$. The expectation value of the Wilson loop is
\be
\label{firstorderWilsonloop}
\b \tr_j W_C\ket = 
\frac{1}{Z}
\int\left({\ts\prod\limits_{x\mu}}\;\d U_{x\mu}\right) \sum_{j_{x\mu\nu}}\;
\left(\prod_{x\mu\nu}\,(2j_{x\mu\nu}+1)\,\chi_{j_{x\mu\nu}}(W_{x\mu\nu})\,\,\e^{-\frac{2}{\beta}\,j_{x\mu\nu}(j_{x\mu\nu} + 1)}\right)
\chi_j(W_C)\,.
\ee
As in eq.\ \eq{plaquetteholonomy}, we express holonomies in terms of Lie algebra elements:
\be
\label{holonomyWilsonloop}
U^{s_i}_{e_i} = \e^{\irm\,\theta^a_i \sigma^a/2}\,,\qquad\mbox{and}\qquad W_C  = \e^{\irm\,\omega^a_C \sigma^a/2}\,.
\ee
By going through the same steps that led to \eq{latticeBFYMPoissonformulaapplied}, we obtain
\bea
\lefteqn{\b \tr_j W_C\ket = \frac{1}{Z}\int\left({\ts\prod\limits_{x\mu\nu}}\;\d U_{x\mu}\right) 
\int_{\bR^3}\left({\ts\prod\limits_{x\mu\nu}}\;\d^3 b_{x\mu}\,\sum_j\,\delta(|b_{x\mu\nu}| - j)\right)} \nonumber \\
&& 
\times\,\frac{(2j+1)}{4\pi}\int_{S^2}\d n\;\frac{|\omega_C|/2}{\sin(|\omega_C|/2)}
\left({\ts\prod\limits_{x\mu\nu}}\,\frac{|\omega_{x\mu\nu}|/2}{\sin(|\omega_{x\mu\nu}|/2)}\right) \nonumber \\
&& \times\,
\exp\left[\sum_x
\left(\frac{\irm}{2}\,b_{x\mu\nu}\cdot\omega_{x\mu\nu} 
- \frac{1}{\beta}\, b^2_{x\mu\nu} + \frac{\irm}{2}\,(2j+1)\,n\cdot\omega_C\right)
\right]\,.
\label{WilsonlooplatticeBFYM}
\eea

\section{Representation as dual gluons and monopole-like excitations}
\label{representationasgluonsandmonopolelikeexcitations}

In U(1) lattice gauge theory confinement is an effect of the compact group topology and cannot be derived within a purely perturbative scheme.
For SU(2) a central question is therefore the following: how can we perform an analytic computation of the quark potential that takes proper account of the compactness of the gauge group?

In the derivation of the photon-monopole representation by Banks et al.\ the group variables are integrated out. After this, the compactness of U(1) resides in the discreteness of the charge variables $l$. By application of the Poisson summation formula, the discreteness is traded in for monopole degrees of freedom. 

In this paper, we want to do something similar: in the lattice BF Yang-Mills theory, the compactness of the gauge group is reflected by the compact range of the group variables and the discrete lengths of the $b$-vectors. One possibility would be to integrate out the group variables as for U(1)---leading to the spin foam representation---and then to attempt a computation. This avenue is pursued in the companion paper II.

Here, we will eliminate the group variables by using a stationary phase argument. In the first step, we apply the Poisson summation formula in expression \eq{WilsonlooplatticeBFYM} to replace the discrete $b$-variable by a continuous $b$ and a discrete monopole-like variable $m$. Then, we apply a stationary phase approximation to the path integral, and thereby obtain a simple constraint on the connection and the $b$-field. The solution to the constraint is plugged back into the path integral, and this gives us the dual gluon degrees of freedom.

\subsection*{Poisson summation formula}

Using the Poisson summation formula, we replace the discrete sum over lengths of $b$ by an integral over lengths and a discrete sum over a new variable $m$:
\bea
\lefteqn{\int_{\bR^3}\d^3 b\,\sum_j\,\delta(|b| - j)
\exp\left[\frac{\irm}{2}\,b\cdot\omega 
- \frac{1}{\beta}\, b^2\right]} 
\nonumber \\
&&= 
\int_0^\infty \d r\int_0^{2\pi}\d\phi\int_0^{\pi}\d\vartheta\;
\sum_j\,\delta(r - j)\;
\exp\left[\frac{\irm}{2}\,r|\omega|\cos\vartheta
- \frac{1}{\beta}\, b^2\right] 
\nonumber \\
&&= 
\frac{1}{2}\;\int_{-\infty}^\infty \d r\;\ldots \sum_{j\in\bZ/2}\,\delta(r - j)\;\ldots \nonumber \\
&&= 
\frac{1}{2}\;\int_{-\infty}^\infty \d r\int_0^{2\pi}\d\phi\int_0^{\pi}\d\vartheta\;
\sum_{m\in\bZ}\;\e^{4\pi\irm\,m\,r}\;
\exp\left[\frac{\irm}{2}\,r|\omega|\cos\vartheta 
- \frac{1}{\beta}\, b^2\right] 
\nonumber \\
&&= 
\int_0^\infty \d r\int_0^{2\pi}\d\phi\int_0^{\pi}\d\vartheta\;
\sum_{m\in\bZ}\;
\exp\left[\frac{\irm}{2}\,r|\omega|\cos\vartheta 
- \frac{1}{\beta}\, b^2 + 4\pi\irm\,m\,r\right] 
\nonumber \\
&&= 
\int_{\bR^3}\d^3 b\;
\sum_{m\in\bZ}\;
\exp\left[\frac{\irm}{2}\,b\cdot\omega 
- \frac{1}{\beta}\, b^2 + 4\pi\irm\,|b| m\right] 
\eea
With this the path integral \eq{WilsonlooplatticeBFYM} becomes
\bea
\b \tr_j W_C\ket &=& \frac{1}{Z}\int\left({\ts\prod\limits_{x\mu}}\;\d U_{x\mu}\right) 
\int_{\bR^3}\left({\ts\prod\limits_{x\mu\nu}}\;\d^3 b_{x\mu\nu}\right) 
\sum_{m_{x\mu\nu}} \nonumber \\
&& 
\times\,\frac{(2j+1)}{4\pi}\int_{S^2}\d n\;\frac{|\omega_C|/2}{\sin(|\omega_C|/2)}
\left({\ts\prod\limits_{x\mu\nu}}\,\frac{|\omega_{x\mu\nu}|/2}{\sin(|\omega_{x\mu\nu}|/2)}\right) \nonumber \\
&& \times\,
\exp\left[\sum_x
\left(\frac{\irm}{2}\,b_{x\mu\nu}\cdot\omega_{x\mu\nu} 
- \frac{1}{\beta}\, b^2_{x\mu\nu} + 4\pi\irm\,|b_{x\mu\nu}| m_{x\mu\nu} + \frac{\irm}{2}\,(2j+1)\,n\cdot\omega_C\right)
\right]\,. 
\label{latticeBFYMPoissonformulaapplied}
\eea
In the term containing $m_{x\mu\nu}$, the repeated indices $\mu$ and $\nu$ are only summed over the pairs $\mu < \nu$.

\subsection*{Stationary phase approximation}

In order to determine the stationary phase point of the path integral, we rewrite the prefactors in expression \eq{latticeBFYMPoissonformulaapplied}
as exponentials. Taylor expansion around $\theta_{x\mu} = \omega_{x\mu\nu} = 0$ yields
\be
\label{expansionofprefactors}
\frac{|\omega_{x\mu\nu}|/2}{\sin(|\omega_{x\mu\nu}|/2)}
= \exp\left(\frac{1}{6}\left(\frac{|\omega_{x\mu\nu}|}{2}\right)^2 + \frac{1}{180}\left(\frac{|\omega_{x\mu\nu}|}{2}\right)^4 + \ldots\right)
\ee
and
\be
\frac{|\omega_C|/2}{\sin(|\omega_C|/2)} = \exp\left(\frac{1}{6}\left(\frac{|\omega_C|}{2}\right)^2 + \frac{1}{180}\left(\frac{|\omega_C|}{2}\right)^4 + \ldots\right)\,.
\ee
The Baker-Campbell-Hausdorff formula for $n$-fold products of group variables \cite{FreidelLouaprediffeomorphisms} gives us
\be
\label{BCHformula}
\omega_C = \sum_i\Big(\theta_i + \left[\Omega_i,\theta_i\right]\Big)\,.
\ee
$\Omega_i$ is a Lie algebra element whose first terms read \setlength{\jot}{0.3cm}
\bea
\Omega_i &=& \hspace{0.5cm}\frac{\irm}{2}\sum_{i<m} \theta_i \\
&& {} - \frac{1}{6}\sum_{m<n<i} [\theta_m,\theta_n] 
+ \frac{1}{6}\sum_{i<m<n} [\theta_m,\theta_n]
- \frac{1}{12}\sum_m [\theta_m,\theta_i] \\
&& {}+\;\;\ldots  \setlength{\jot}{0cm}
\eea
Further terms are given by higher order commutators. Let us define the source current
\be
\label{source}
J = (j+1/2)\,n\;C\,
\ee
so that
\be
(j+1/2)\,n\cdot\sum_i \theta_i = \sum_x J_{x\mu}\!\cdot\theta_{x\mu}\,.
\ee
Then, after collecting all factors, the total exponent is
\bea
\clS &=& 
\sum_x
\Big(- \frac{1}{\beta}\, b^2_{x\mu\nu} + 4\pi\irm\,|b_{x\mu\nu}| m_{x\mu\nu} \nonumber \\
&& \hspace{0.85cm}{}- \frac{\irm}{2}\,b_{x\mu\nu}\cdot\left(\nabla_\mu\theta_{x\nu} - \nabla_\nu\theta_{x\mu} + \theta_{x\mu}\times \theta_{x\nu} + \ldots\right) \nonumber \\
&& \hspace{0.85cm}{}- \irm\,J_{x\mu}\!\cdot\theta_{x\mu} + \ldots \nonumber \\
&& \hspace{0.85cm}{}+ \frac{1}{24}\,\omega_{x\mu\nu}^2 + \ldots \nonumber \\
&& \hspace{0.85cm}{}+ \frac{1}{24}\,\omega_C^2 + \ldots\Big)\,,
\eea
where dots indicate the higher order terms in the respective expansions. Variation w.r.t.\ $\theta_{x\mu}$ and $b_{x\mu\nu}$ leads to
\be
\label{firstcondition}
\nablab_\nu b_{x\mu\nu} = J_{x\mu} + \theta_{x\mu}\times b_{x\mu\nu} + \ldots
\ee
and
\be
\label{secondcondition}
-\frac{2}{\beta}\,b_{x\mu\nu} 
+ 4\pi\irm\,\frac{b_{x\mu\nu}}{|b_{x\mu\nu}|}\,m_{x\mu\nu} 
- \frac{\irm}{2}\left(\nabla_\mu\theta_{x\nu} - \nabla_\nu\theta_{x\mu} + \theta_{x\mu}\times \theta_{x\nu} + \ldots\right) = 0\,.
\ee
For Fourier modes $\theta_{x\nu}(k)$ where $k$ is sufficiently large, the last equation is dominated by the derivative terms, so we get a solution $(\theta,b)$ by setting $\omega_{x\mu\nu} = 0$ and solving \eq{firstcondition}. If we ignore non-trivial configurations due to the topology of the periodic lattice, this solution is gauge-equivalent to $(\theta,b)$ where $\theta_{x\mu} = 0$ and $b$ is a solution of
\be
\label{abelianGaussconstraint}
\nablab_\nu b_{x\mu\nu} = J_{x\mu}\,.
\ee
In abstract notation, this means that
\be
\pa b = J\,.
\ee
It has the general solution
\be
b = \pa\!*\!\varphi + \bb\,,
\ee
or in index notation
\be
\label{generalsolution}
b_{x\mu\nu} = \epsilon_{\mu\nu\rho\sigma_1\cdots\sigma_{d-3}}\nablab_\rho\varphi_{x\sigma_1\cdots\sigma_{d-3}} + \bb_{x\mu\nu}\,.
\ee
Here, $\varphi$ is an $\bR^3$-valued $(d-3)$-chain on the dual lattice $\kappa^*$ and $\bb$ is a particular solution of the inhomogeneous equation. We fix the latter as
\be
\bb_{x\mu\nu} = -u_\mu\,(u\cdot\nabla)^{-1} J_{x\nu} + u_\nu\,(u\cdot\nabla)^{-1} J_{x\mu}\,.
\ee
For sufficiently low momenta $k$, on the other hand, the derivative terms in \eq{secondcondition} can be ignored, and we deal instead with the two equations \setlength{\jot}{0.4cm}
\bea
&& \nablab_\nu b_{x\mu\nu} = J_{x\mu} + \theta_{x\mu}\times b_{x\mu\nu} + \ldots\,, \\
&& -\frac{2}{\beta}\,b_{x\mu\nu} 
+ 4\pi\irm\,\frac{b_{x\mu\nu}}{|b_{x\mu\nu}|}\,m_{x\mu\nu} 
- \frac{\irm}{2}\,\theta_{x\mu}\times \theta_{x\nu} = 0\,.
\eea
In that case, the higher orders in the connection and curvature do matter, since the curvature can be large. 
\setlength{\jot}{0cm}

The crucial step in our derivation is the following: we will proceed \textit{as if} the solution for the fast modes could be used for the whole momentum range: that is, we will take the solution $(\theta,b) = (0,\pa *\!\varphi + \bb)$ and its gauge-transforms, and restrict the path-integral to these ``stationary phase points''. If the current and monopoles were set to zero, this would correspond to the zeroth order in weak-coupling perturbation theory of BF Yang-Mills theory (see sec.\ 2.5 of \cite{Cattaneoetal}). In this sense, our method is perturbative. When current and monopoles are switched on, however, it becomes a perturbation theory around non-trivial field configurations, and thereby includes non-perturbative effects. The non-trivial configurations appear in two ways: 1.\ Similarly as in the photon-monopole description of U(1) lattice gauge theory, we obtain monopole-like variables that preserve information about compactness. 2.\ As part of the solution to \eq{abelianGaussconstraint}, we get the particular solution $\bb$ which carries large-distance information about the ``defect'' created by the source $J$. 

We now restrict the path integral \eq{latticeBFYMPoissonformulaapplied} to the above solutions, i.e.\ to flat configurations given by $(\theta,b) = (0,\pa\!*\!\varphi + \bb)$ and its gauge-transforms. In doing so, we observe the following: 
1.\ The integration over gauge-transformations can be factored off and is trivial, since the Haar measure is normalized. Thus, it suffices to integrate over all configurations of the type $(\theta,b) = (0,\pa\!*\!\varphi + \bb)$.
2.\ The constrained $b$-integral can be represented as an integral over $\varphi$. Since the map from scalar field to $b$-field is degenerate, however, we need to introduce a ``gauge-fixing'' on $\varphi$. 3.\ The Jacobian for the change of variables $b\to \varphi$ is a constant and may be omitted in expectation values. 
4.\ As part of the stationary phase approximation, we also receive factors from the integration over the quadratic order in fluctuation variables.
We will ignore this contribution.

\subsection*{Three dimensions}

In 3 dimensions equation \eq{generalsolution} reads
\be
\label{generalsolution3d}
b_{x\mu\nu} = \epsilon_{\mu\nu\rho}\nablab_\rho\varphi_x + \bb_{x\mu\nu}
\ee
Up to a constant, $\varphi$ is determined by $b$ and $\bb$, so the degeneracy appears only for the mode of zero momentum.
To avoid an overcounting, we remove the integration over this zero mode from the $\varphi$-integral. The resulting path integral reads
\bea
\lefteqn{\b \tr_j W_C\ket =
\frac{1}{Z}\int_{\bR^3}\left({\ts\prod\limits_x}\;\d^3\varphi_x\right)'\sum_{m_{x\mu}}\;\frac{(2j+1)}{4\pi}\int_{S^2}\d n} \nonumber \\
&& \times\,
\exp\left[\sum_x
\left(-\frac{2}{\beta}
\left(\nablab_\mu\varphi_x + \bb_{x\mu}\right)^2 + 4\pi\irm\left|\nablab_\mu\varphi_x + \bb_{x\mu}\right|m_{x\mu}
\right)\right]
\eea
For notational convenience we switched from 2-chains $\bb$ and $m$ to 1-chains on $\kappa^*$, i.e.
\be
\bb_{x\rho} = \frac{1}{2}\,\epsilon_{\rho\mu\nu} \bb_{x\mu\nu}\,,\qquad m_{x\rho} = \frac{1}{2}\,\epsilon_{\rho\mu\nu} m_{x\mu\nu}\,.
\ee
It is understood that the same stationary phase approximation is applied to the partition function, i.e.\ $Z$ is the same path integral with $\bb$ set to zero.

We can rewrite this expression further by facoring off a short-distance potential: by a change of variables 
\be
\varphi_x - \Delta^{-1}\nabla_\mu\bb_{x\mu}\quad\rightarrow\quad\varphi_x\,,
\ee
and using the identity 
\be
\label{identityforJSU2}
\nabla_\mu\bb^a_{x\mu}\Delta^{-1}\nabla_\mu\bb^a_{x\mu} + \bb^2_{x\mu} = - J^a_{x\mu}\Delta^{-1}J^a_{x\mu}\,,
\ee
we extract a $1/k^2$-potential:
\bea
\lefteqn{\b \tr_j W_C\ket =
\frac{1}{Z}\int_{\bR^3}\left({\ts\prod\limits_x}\;\d^3\varphi_x\right)'\sum_{m_{x\mu}}\;\frac{(2j+1)}{4\pi}\int_{S^2}\d n} \nonumber \\
&& \times\,
\exp\left[\sum_x
\left(\frac{2}{\beta}\,\varphi^a_x\Delta\varphi^a_x 
+ 4\pi\irm\left|\nablab_\mu\left(\varphi_x + \Delta^{-1}\nabla_\nu\bb_{x\nu}\right) + \bb_{x\mu}\right|m_{x\mu}
+ \frac{2}{\beta}\,J^a_{x\mu}\Delta^{-1}J^a_{x\mu}
\right)\right]\,. \nonumber \\
\label{result3d}
\eea

\subsection*{Four dimensions}

In $d=4$ equation \eq{generalsolution} takes the form
\be
\label{generalsolution4d}
b_{x\mu\nu} = \epsilon_{\mu\nu\rho\sigma}\nablab_\rho\varphi_{x\sigma} + \bb_{x\mu\nu}\,.
\ee
We can consider $\varphi_{x\mu}$ as a gauge potential for a dual field strength
\be
\nablab_\mu\varphi_{x\nu} - \nablab_\nu\varphi_{x\mu}\,. 
\ee
$\varphi_{x\mu}$ is not uniquely determined by $b_{x\mu\nu}$ and $\bb_{x\mu\nu}$. To remove the ambiguity, we impose a gauge condition, say, the axial gauge
\be
\varphi_{x1} = 0\,.
\ee
Then, the path integral takes the form
\bea
\lefteqn{\b \tr_j W_C\ket = 
\frac{1}{Z}\int_{\bR^3}\left({\ts\prod\limits_{xi}}\;\d^3\varphi_{xi}\right)'\sum_{m_{x\mu\nu}}\;\frac{(2j+1)}{4\pi}\int_{S^2}\d n} \nonumber \\
&& \times\,
\exp\left[\sum_x
\left(-\frac{1}{\beta}
\left(\epsilon_{\mu\nu\rho i}\nablab_\rho\varphi_{x i} + \bb_{x\mu\nu}\right)^2 
+ 4\pi\irm\left|\epsilon_{\mu\nu\rho i}\nablab_\rho\varphi_{x i} + \bb_{x\mu\nu}\right|m_{x\mu\nu}
\right)\right]\,.
\eea
The Roman index $i$ takes the values 2, 3, 4. In the last term of the exponent the indices $\mu$ and $\nu$ are only summed over pairs $\mu<\nu$.

Again, we factor off a $1/k^2$ potential: this time we use that for $u = \oneh$
\be
\frac{1}{4}\left(\epsilon_{i\rho\mu\nu}\nabla_\rho\bb^a_{x\mu\nu}\right)
\Delta^{-1}
\left(\epsilon_{i\sigma\kappa\lambda}\nabla_\sigma\bb^a_{x\kappa\lambda}\right) 
+ \frac{1}{2}\,\bb_{x\mu\nu}^2 = - J^a_{x\mu}\Delta^{-1}J^a_{x\mu}\,,
\ee 
and get
\bea
\lefteqn{
\b \tr_j W_C\ket = 
\frac{1}{Z}\int_{\bR^3}\left({\ts\prod\limits_{xi}}\;\d^3\varphi_{xi}\right)'\sum_{m_{x\mu\nu}}\;\frac{(2j+1)}{4\pi}\int_{S^2}\d n} \nonumber \\
&& \hspace{-1cm}\times\,\exp\left[\sum_x\left(
\frac{2}{\beta}\,\varphi^a_{xi}\Delta\varphi^a_{xi} 
+ \frac{2}{\beta}\,\left(\nabla_i \varphi_{xi}\right)^2 
+ 4\pi\irm\left|\epsilon_{\mu\nu\rho i}\nablab_\rho\left(\varphi_{x i} 
+ \epsilon_{i\sigma\kappa\lambda}\,u_\sigma\Delta^{-1}\nabla_\kappa J_{x\lambda}\right) + \bb_{x\mu\nu}\right|m_{x\mu\nu} \right.\right. \nonumber \\
&& \hspace{1.2cm} {}+ \frac{2}{\beta}\,J^a_{x\mu}\Delta^{-1}J^a_{x\mu}
\bigg)\Bigg]\,. \label{result4d}
\eea

\subsection*{Dual gluons and monopole-like excitations}

We propose \eq{result3d} and \eq{result4d} as a non-abelian generalization of the photon-monopole representation of U(1) lattice gauge theory. 
The field $\varphi$ is interpreted as a dual gluon field, similar to the dual photon field of the abelian case:
\begin{itemize}
\item $\varphi$ mediates the short-distance interaction 
\be
V_{JJ} := -\frac{2}{\beta}\,\sum_x\;J^a_{x\mu}\Delta^{-1}J^a_{x\mu} = -\frac{1}{2}\,a g^2 (j+1/2)^2\sum_{xy}\;C_{x\mu}\Delta^{-1}_{xy} C_{y\mu}\,,\qquad j\neq 0\,.
\ee
The latter agrees roughly\footnote{The reader may wonder why the formula gives a nonzero potential when $j$ is zero. The answer is that we \textit{do} get a zero potential when the spin is zero from the start. If we use the Kirillov trace formula, however, and set $j=0$ at the end of the derivation, the stationary phase approximation creates an error and a nonzero offset in the $j$-dependence.} 
with the tree-level result of standard perturbation theory: there one would have \cite{Petertothreeloop,Peterfulltwoloop}
\be
V^{\mathrm{tree}}_{JJ} = -\frac{1}{2}\,a g^2 j(j+1)\sum_{xy}\; C_{x\mu}\Delta^{-1}_{xy} C_{y\mu}\,.
\ee
\item $\varphi$ has 3 degrees of freedom per point, which agrees with the fact that in 3 dimensions we have 1 physical degree of freedom per gluon and altogether 3 gluons for SU(2).
\end{itemize}
The $m$-variables are reminiscent of the discrete monopole variables of U(1) lattice gauge theory.
We refer to them as monopole-like excitations. The analogy with U(1) is not complete, however, since 
in our case the $m$'s are $(d-2)$-chains on the dual lattice, while they are $(d-3)$-chains for U(1).

\section{Summary and discussion}
\label{summaryanddiscussion}

In this paper, we have derived an approximative representation for SU(2) lattice gauge theory in dimension 3 and 4. Its degrees of freedom can be seen as 
a dual gluon field and a monopole-like field. We propose it as a generalization of the photon-monopole representation of Polyakov \cite{PolyakovI,PolyakovII} and Banks, Myerson and Kogut \cite{BanksMyersonKogut}.

We started by rewriting the SU(2) lattice Yang-Mills theory as a BF Yang-Mills theory on the lattice. Then, we transformed the expectation value of a Wilson loop in several steps to a path integral over a dual gluon field and monopole-like variables. This was done by using the Poisson summation formula, a stationary phase approximation and by solving a Gauss constraint. 

The critical step of the derivation is the stationary phase approximation. It can be viewed as the zeroth order of a weak-coupling perturbation theory that has two types of non-trivial field configurations as a background: 1.\ the monopole-like excitations that arise from the compactness of the gauge group, and 2.\ the particular solution of the Gauss constraint which carries information about the large-distance defect created by the current.

The resulting model contains two interaction terms: firstly, a current-current potential that is essentially the tree-level Coulomb interaction one would get from a purely perturbative treatment. Secondly, a coupling between monopole-like excitations, dual gluons and current. This coupling is similar to the photon-monopole coupling of U(1), but nonlinear.  

In the two companion papers, we start from two other representations of 3d SU(2) lattice gauge theory and arrive at models that are quite similar to the present one. The similarity to the photon-monopole representation of U(1) suggests that these models could have interesting large-distance properties, i.e.\ exhibit confinement.

Whether the model of this paper is a good or a bad approximation can be tested: it is simpler than the full lattice gauge theory and may be easily implemented on a computer. By summing over the monopole variables in expression \eq{result3d} one can remove the phase factors and translate them into a constraint. The latter can be enforced by a Gaussian damping factor.

\section*{Acknowledgements}

I thank Abhay Ashtekar, Gerhard Mack, Alejandro Perez and Hendryk Pfeiffer for discussions. 
This work was supported in part by the NSF grant PHY-0456913 and the Eberly research funds.

\bibliography{bibliography}
\bibliographystyle{hunsrt}  

\end{document}